\documentclass[aps,prb,twocolumn,superscriptaddress]{revtex4}
\usepackage{graphicx}
\usepackage{bm}

\begin{document}

%\preprint{}

\title{
Novel Microwave Absorption Due to Strong Coupling between
Josephson Plasma and the Josephson Vortex Array
in Bi$_2$Sr$_2$CaCu$_2$O$_{8+\delta}$}



\author{I. Kakeya}
\email[]{kakeya@ims.tsukuba.ac.jp}
%\thanks{}
\affiliation{Institute of Materials Science, University of Tsukuba, Tsukuba, Ibaraki 305-8573 Japan}
\author{T. Wada}
\affiliation{Institute of Materials Science, University of Tsukuba, Tsukuba, Ibaraki 305-8573 Japan}
\author{K. Kadowaki}
\affiliation{Institute of Materials Science, University of Tsukuba, Tsukuba, Ibaraki 305-8573 Japan}


\date{\today}
\begin{abstract}
%\baselineskip=40pt
We have investigated the Josephson plasma excitations 
in magnetic fields parallel 
to the $ab$-plane in Bi$_2$Sr$_2$CaCu$_2$O$_{8+\delta}$ single crystals
in a wide microwave frequency region (9.8 -- 75 GHz).
It was found that there are two kind of phase-collective modes: 
one increases with magnetic fields in higher fields,
approaching linear asymptotic dependence with $ck$,
which lies well above the inherent plasma frequency $\omega_p$,
while the other does not show considerable field dependence.
The higher linear mode is attributed to the Josephson plasma mode 
propagating along the reciprocal lattice vector of 
the Josephson vortex,
whereas the lower one can be ascribed to the oscillation mode of 
Josephson vortices.

\end{abstract}

%\keywords{}

\maketitle

%\section{}
The interplay between the Josephson plasma (JP) and the Josephson vortex (JV),
which is due to the linear and non-linear phase dynamics of 
the Josephson junction,
has attracted much attention since the 1960s.
In single junctions, the excitation spectrum has been investigated
by Fetter and Stephen~\cite{Fet68}.
They derived a gapless ``vortex'' mode attributed to the dc sliding 
of JVs 
and a ``plasma'' mode identical to the transverse JP, 
in which the propagation vector $\bm{k}$ is equal to the primitive reciprocal
lattice vector of the one-dimensional JV array.
The frequency of the traveling plasma wave can be controlled
by the field parallel to the block (barrier) layer $H_{\parallel}$
according to the dispersion relation of the transverse plasma
$\omega_p^T=\omega_p\sqrt{1+(ck)^2}$, 
where $\omega_p\equiv c/\sqrt{\epsilon}\lambda_c$ is the inherent plasma frequency,
with $\epsilon$ and $\lambda_c$ being the high-frequency dielectric constant 
and the penetration depth along the $c$ axis.
This plasma mode explains the so-called Ech resonance,
which was found in the current--voltage ($I-V$) characteristics 
of Pb/PbO/Pb junctions,
with the resonance voltage being proportional to $H_{\parallel}$\cite{Eck64}.

In systems where the atomic-scale of weak Josephson junctions are stacking,
namely an intrinsic Josephson junction (IJJ),
which is realized in anisotropic high-$T_c$ superconductors
represented by Bi$_2$Sr$_2$CaCu$_2$O$_{8+\delta}$ (BSCCO)~\cite{Kle94},
the physics attributed to JP, JV, and their interplay 
are drastically richer\cite{Tac94,Cle90}.
Because of the larger charge screening length ($\mu \simeq$ 10 \AA) 
compared with the thickness of the CuO$_2$ bilayers ($t=3$ \AA),
Josephson coupling between not only adjacent layers but also 
next-adjacent layers contributes to the charge transport along the $c$ axis
\cite{Mac99}.
This longitudinal coupling yields the longitudinal JP mode 
propagating along the $c$ axis with a dispersion relation of 
$\omega_p^L=\omega_p \sqrt{1+\epsilon \mu^2}$
and rich variety of the dynamical nature of the two-dimensional JV lattice,
which has been predicted by theoretical calculations\cite{Mac00} 
and partly found experimentally\cite{Hec97}.

So far, studies on the JP of high-$T_c$ superconductors 
in finite magnetic fields
have concentrated on the case of 
the $c$ axis field\cite{Shi99a,plasma2ka},
where vortices perpendicular to the layers can be regarded as
static matter to suppress the Josephson coupling,
representing fluctuations of pancake vortices\cite{Kos96}.
In contrast, in the field parallel to the $ab$ plane,
vortices should be treated as dynamical matter because of the strong coupling 
between JP and JV.
The microwave absorption associated with JP in the presence of JV 
is one of the key phenomena in revealing the linear and nonlinear 
dynamics of IJJ,
which may also bring about high-frequency applications of high-$T_c$ materials. 

In this letter,
we describe the field dependence of the plasma frequency 
$\omega_p(H_{\parallel})$
in fields parallel to the $ab$-plane through the Josephson plasma resonance 
(JPR) measurements 
as functions of the microwave frequency, field intensity, and temperature,
in order to reveal the entire nature of JP in 
the presence of JVs.
We obtained two resonance modes 
at higher and lower frequencies 
in good accordance with recent theoretical results.
The higher-frequency mode, 
which increases in proportion 
with the field, is attributed to JP mode
propagating along the primitive reciprocal lattice vector of the JV lattice, 
while the lower frequency mode, which depends only weakly on the field  
is interpreted as the vortex oscillation mode 
modified by the vortex pinning and stacking effects of IJJ.
This result explains the preliminary results of Matsuda {\it et al}.,
who have reported vanishing of the plasma resonance in fields very close to the $ab$ plane\cite{Mat97a}.

We have measured JPR in three BSCCO single crystals grown 
by the modified traveling solvent floating zone method. 
Two of the crystals were under-doped (U1, U2), 
while the other was optimallydoped (OP), 
with all having been
obtained after appropriate annealing processes in controlled atmospheres.
The superconducting transition temperatures $T_c$'s of these crystals were 
determined by magnetization measurements with a 
SQUID magnetometer as 70.2, 72.5, and 90.5 K for U1, U2, and OP, respectively. 
All JPR data displayed here were obtained in U1.
JPR measurements were made in a microwave frequency range between 
9.8 and 75 GHz, with both reflective and transmission-type
bridge balance circuits 
with rectangular TE$_{102}$ mode cavity resonators.
To selectively excite the longitudinal plasma mode only, 
the sample was placed inside the cavity 
so that the oscillating electric field parallel to the $c$-axis 
was exerted homogeneously over the $ab$-plane of the crystal~\cite{PRB98}.
Frequency-stabilized microwaves were generated by the signal swept generator 
(HP83650B) or Gunn oscillators,  
and the resonance was detected by changes in the cavity impedance (Q-factor) 
by sweeping temperatures. 
The magnetic field was applied by a split-pair superconducting magnet,
and the angle between field direction and the CuO$_2$ plane was adjusted by
rotating the cavity resonator with respect to the magnet 
by a precision rotator.
The alignment of the field direction exactly parallel to the $ab$-plane 
was determined from
the symmetry of the angular dependence of the resonance within 
an accuracy of 0.01 degrees.

\begin{figure}
\begin{center}
\includegraphics[width=\linewidth]{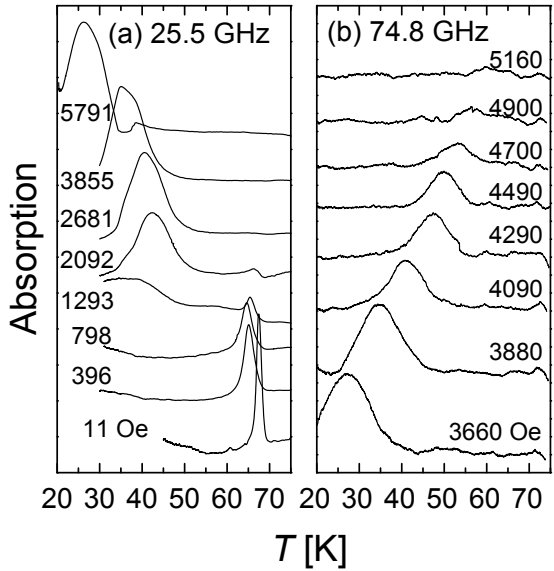}
\end{center}
\caption{
JPR absorption curves obtained by sweeping temperature at 
(a) 25.5 and (b) 74.8 GHz at various magnetic fields
parallel to the $ab$-plane. 
Temperature was swept from far above $T_c$ 
to the lowest temperature 
after setting a constant magnetic field, and no hysteresis was found.
}
\label{fig:rawdata1-1}
\end{figure}

Figures \ref{fig:rawdata1-1} (a) shows the JPR absorption
 curves at 25.5 GHz 
obtained in various constant fields by sweeping temperature.
Two clear resonance lines without hysteresis are found at higher and lower temperatures
with a temperature gap in between.
A sharp and symmetric resonance was observed just below
$T_c$ in a low field region.
With increasing fields, the higher temperature resonance mode shifts to 
lower temperatures below 1 kOe, then turns back to higher temperatures,
and finally disappears above 1.5 kOe.
In contrast, 
the resonance found at lower temperatures appears above 1.2 kOe.
The resonance shifts to lower temperatures 
without a considerable change in its line-shape and intensity 
as the field is increased.
The measurements with sweeping temperature at constant field 
were ideally reversible with well-defined low-resonance peaks at
the higher temperatures and rather broad higher-field resonances at 
lower temperatures.
We refer to the higher-temperature and lower-temperature
resonances as HTB (higher temperature branch) and LTB peaks, respectively.
That these two resonance branches are observed simulatanously 
and have different field dependences 
is peculiar to JPR in $\bm{H} \parallel ab$ 
and is in sharp contrast to JPR in $\bm{H} \parallel c$,
where one resonance line whose resonance field monotonically decrease 
with temperature is observed in the configuration of the longitudinal
excitation\cite{PRB98}.

\begin{figure}[tb]
\begin{center}
\includegraphics[width=0.9\linewidth]{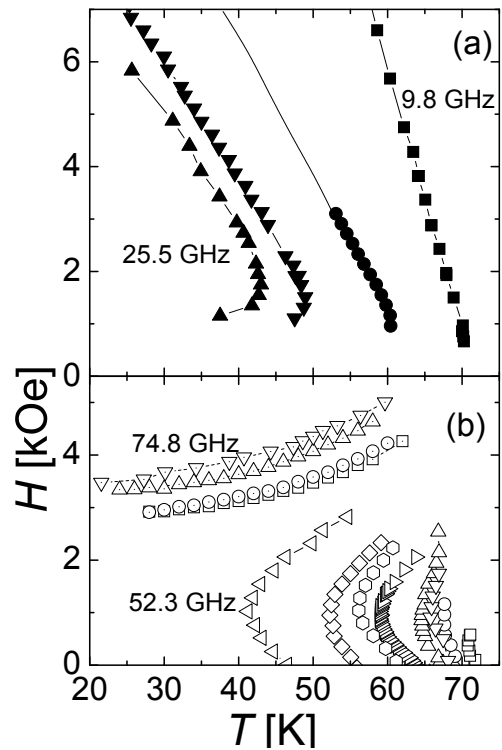}
\end{center}
\caption{
Field--temperature diagrams of the resonance modes for various frequencies.
(a): LTB obtained at 9.8, 18.8, 22.3, and 25.5 GHz. 
(b): HTB obtained at 9.8, 18.8, 22.3, 25.5, 34.5, 39.5, 44.2, 52.3 61.7, 65.9, 74.3, and 74.8 GHz from right to left.
Symbols with the same contours represent data at the same frequencies.
}
\label{fig:plot1-2}
\end{figure}

To study the doping dependence of these features, 
we performed similar measurements in crystals OP and U2. 
The JPR results at $\omega/2\pi=$ 42 GHz in OP 
were found to be qualitatively similar to the ones at 19 GHz in U1, 
although the resonance temperatures and fields were different.
This similarity can be understood by considering
the frequency ratio of $\omega/\omega_p$,
which is less than one and closer to unity 
in more under-doped crystals for a given $\omega$ 
because $\omega_p$ is lower in higher anisotropic crystals.
Provided that 19 GHz in U1 and 42 GHz in OP 
give the same $\omega/\omega_p$ value,
we derive $\omega_p$ of OP as 130 GHz, 
which is a typical plasma frequency for optimallydoped BSCCO.
Here, we used $\omega_p$ of U1 as 56.8 GHz, as described below.
This scaling is also valid for results in U2, 
with slightly higher doping than U1.
Therefore, 
the resonance splitting is observed at $\omega \gtrsim \omega_p/3$, 
and the magnitude of the temperature gap between LTB and HTB
is larger at $\omega$, being closer to $\omega_p$.
In contrast, $\omega$ is much smaller than $\omega_p$,
LTB merges to HTB as observed at 9.8 GHz in U1, 
resulting in a single branch.
This is the reason why Matsuda {\it et al}. 
did not observe vanishing of the resonance in an optimallydoped crystal 
in Ref~\cite{Mat97a}.

The resonance peaks of LTB and HTB as a function of magnetic fields
for various frequencies are plotted in Fig. \ref{fig:plot1-2} (a) and (b),
respectively.
With increasing frequency, both LTB and HTB shift to lower temperatures.
Because the decrease of LTB is larger than the one of HTB, 
LTB goes above our experimental temperature range above 30 GHz,
whereas HTB could be observed throughout the temperature range below 60 GHz.
The JPR can be clearly observed even in zero magnetic field,
as seen in Fig. \ref{fig:rawdata1-1}(a).
This zero-field resonance occurs at a particular temperature $T_0$ and 
at a corresponding microwave frequency.
The frequency dependence of $T_0$ has been reported previously and has been explained by 
using a simple two-fluid model with 
a temperature-independent scattering rate~\cite{New3SC99}.
By extrapolating the temperature dependence of the plasma frequency 
$\omega_p(T)$ to $T=0$, 
the inherent plasma frequency $\omega_p$ is estimated to be 56.8 GHz,
which yeilds $\lambda_c = 217 \mu$m with $\epsilon=15$.

The bending-over behavior of HTB becomes greater at higher frequencies,
and the JPR above 30 GHz extends to 
temperatures even higher than $T_0$.
This surprising behavior is certainly beyond 
our current understandings of JPR in perpendicular fields, 
where the plasma frequency {\it should} decrease 
both with application of high magnetic field and 
with increasing temperature 
because of the suppression of the interlayer coherence 
$\langle \cos \varphi_{l,l+1} (H,T) \rangle$ 
in response to the fluctuation of pancake vortices.
Above 1.5 kOe, HTB obtained at higher frequencies lies at higher fields
at the given temperatures,
suggesting that the HTB resonance can be observed even above 
$\omega_p$.
As shown in Fig. \ref{fig:rawdata1-1} (b),
at 74.8 GHz where no JPR can be observed in the case of $\bm{H}\parallel c$, 
we detected a clear resonance.
This resonance shows a remarkable angular variation and therefore can be
found only within $\pm$ 3 degrees of the $ab$-plane above 3 kOe.
The resonance field of this JPR mode shifts to high fields 
as the temperature is increased and finally disappears above approximately 5 kOe.
A similar resonance was obtained at another three frequencies above 
$\omega_p$ (61.7, 65.9, and 74.3 GHz).
These are plotted in Fig. \ref{fig:plot1-2} (b).

\begin{figure}
\begin{center}
\includegraphics[width=\linewidth]{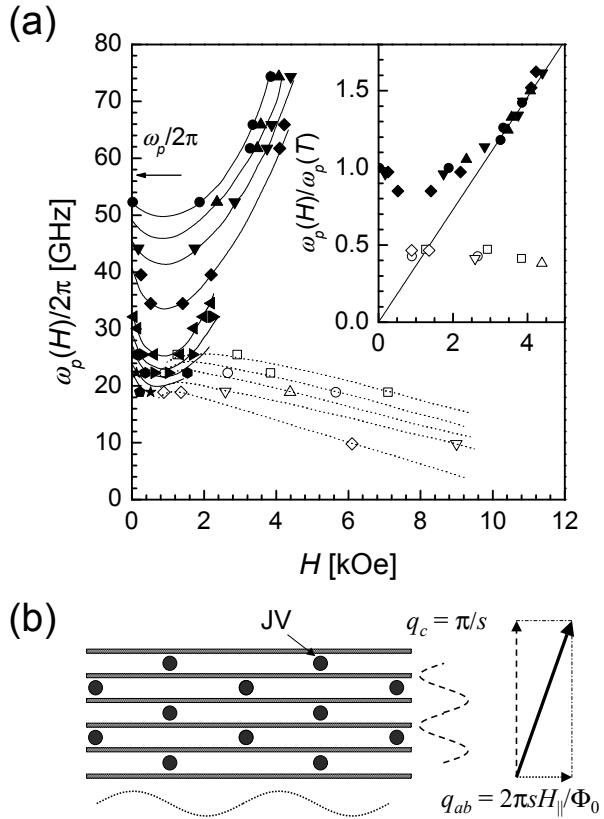}
\end{center}
\caption{
(a) 
The frequency-field diagram for HTB and LTB.
Solid symbols denote HTB at 45, 50, 55, 60, 65, 66, 67, 68, and 69 K, 
and open symbols denote LTB at 40, 45, 50, 55, and 60K from top to bottom.
Solid and dotted curves are guides for the eyes.
In the inset, $\omega_p(H_{\parallel})/\omega_p(T)$ of HTB and LTB 
at 40, 45, 50, 55, and 60 K are shown.
Symbols are identical to those in the main panel, and 
a solid line is given by Eq. (\ref{ParallelPlasma}) with $\gamma=1070$.
(b)
Schematic picture of propagation of the tilted plasma wave.
Interlayer phase difference along the $ab$ plane and the $c$ axis are 
spatially modified as shown by dotted and broken waves, respectively.
The propagation vector of the plasma wave is depicted as a thick arrow
consisting of the transverse component $q_{ab}$ and 
the longitudinal component $q_c$.
}
\label{fig:plot4}
\end{figure}

In Fig. \ref{fig:plot4}(a),
the frequency-field diagrams 
extracted from Fig. \ref{fig:plot1-2} are shown.
It is clear that there are two resonance modes 
well-separated by a frequency gap. 
The higher one extracted from HTB shows an upturn at approximately 1 kOe, 
then monotonically rises to higher frequencies even above 
$\omega_p$, 
while the lower one extracted from LTB shows a gradual decrease with 
increasing field.
Because LTB can be observed only in a finite field, 
the existence of JVs is considered to be required for LTB.
At low fields below 0.5 kOe, 
the extrapolation of LTB to low fields tends to approach zero frequency, 
suggesting that LTB may be in a gapless mode at the zero-field limit.
From these considerations we attribute 
LTB and HTB to the vortex and plasma branches 
as suggested in Ref.~\cite{Fet68}. 

Recently, Bulaevskii {\it et al}. has proposed
a qualitative picture of the plasma resonance in a layered superconductor
on the basis of the single junction model~\cite{Bul97}.
With increasing parallel field $H_{\parallel}$,
the frequency of the plasma resonance increases 
while the resonance intensity decreases because of the decrease in the optical 
weight of the plasma,
and the resonance finally disappears in the high field limit 
$H_{\parallel} \gg H_0 \equiv \Phi_0/\gamma s^2$,
where $\Phi_0$ is the flux quantum and $s$ is the interlayer spacing.
In contrast, 
the vortex resonance survives even in $H_{\parallel} \gg H_0$
and lies at a non-zero frequency by introducing periodic pinning of JVs.
Thus, we attribute HTB and LTB to 
the plasma oscillation and the oscillation of JVs.

Quite recently, 
Koshelev and Machida have derived absorption spectra 
of the JP mode in a layered superconductor 
in high parallel fields, 
with a uniform oscillating electric field being applied 
parallel to the $c$-axis~\cite{Mac01}.
Assuming that all block layers are fully occupied by JVs 
forming a distorted triangular lattice according to the anisotropy,
they analytically and numerically calculated 
the dispersion relation and absorption spectra of the JP mode. 
Because the periodicity of the JV lattice induces
a modulation of the interlayer phase difference, 
the $\bm{k}$-vector of the JP mode corresponds to 
the primitive reciprocal lattice vector of the JV lattice $\bm{q}$,
which consists of the $c$-axis component $q_c$ fixed on $\pi/s$ 
due to the intrinsic pinning 
and the $ab$-plane component $q_{ab}$ proportional to 
the magnetic field $H_{\parallel}$ as 
$2\pi s H_{\parallel}/\Phi_0$, as depicted in Fig. \ref{fig:plot4} (b).
Therefore,
a tilted JP wave 
as a mixture of the longitudinal plasma with $k_z=q_c$ 
and the transverse plasma with $k_{xy}=q_{ab}$ can be excited. 
The plasma frequency is determined by the dispersion relation along $\bm{q}$ 
lying between the longitudinal and transverse plasma modes 
and rises with magnetic field because of the linear increase in $q_{ab}$.
It should be noted that the dispersion is very close to the longitudinal dispersion 
for $q_{ab} \ll q_c$.
As a result, the peak frequency of the dissipation spectrum, 
corresponding to the resonance peak, is proportional to
$H_{\parallel}$ in high fields as 
\begin{equation}
\omega_p(H_{\parallel})=
\omega_p \frac{\pi H_{\parallel} \gamma s^2}{\Phi_0}.
\label{ParallelPlasma}
\end{equation}
Assuming that $\gamma=1070$, 
a value that is quite reasonable for under-doped BSCCO samples, 
this equation provides a good agreement with 
the experimental data above 3 kOe at all temperatures,
as shown in the inset of Fig. \ref{fig:plot4}(a),
where we used $s=15$ {\rm \AA}.
It should be noted that this equation has been derived in Ref.~\cite{Bul97}.
In this situation, 
all block layers are expected to be occupied by JVs above 
$H^*=\Phi_0/4.6\gamma s^2$~\cite{Ich95},
which is estimated to be 2.1 kOe for $\gamma=1070$.
This value is consistent with the fact that the linear field dependence of HTB 
is violated below 3 kOe, as shown in Fig. \ref{fig:plot4}(a).
We therefore conclude that
the linear increase in $\omega_p(H_{\parallel})$ of 
HTB is due to the increase in 
$q_{ab}$ of the JV lattice, 
while $q_c$ is fixed on $\pi/s$ because of the triangular lattice.
This result suggests that the excited plasma frequency can be easily controlled 
by adjusting the magnetic fields parallel to the layers
even above $\omega_p$.

The absorption intensity of the HTB resonance is thought to be 
diminished as frequency is increased
because the intensity is weaker at higher temperature and higher frequency.
This diminishing is predicted due to the increase in the quasiparticle damping 
along the $c$-axis,
which is directly connected to the tunneling mechanism of high-$T_c$ 
superconductors.
As discussed previously~\cite{Tac94}, 
the higher quasiparticle conductivity along the $c$-axis broadens 
the resonance line and weakens the resonance intensity.

The origin of the intensity of the vortex resonance (LTB), 
which is comparable to the plasma resonance in our experiments, 
has not been conclusive, although a much weaker absorption
below $\omega_p$ was obtained by numerical calculations 
in a single junction model in which the randomness
of the critical Josephson current along the layer was introduced~\cite{ISS01}.
The frequency obtained in the calculations is much lower than
the experimental results.
This discrepancy can be attributed to the pancake pinning yielded by
misalignments
of magnetic fields and thermal fluctuations as predicted by Sonin~\cite{Son01}.

The initial decrease in HTB frequency with increasing field below 3 kOe
cannot be explained by their approach, 
which may in part be due to the thermal fluctuation of JVs 
because the slope is larger at higher temperatures.
Furthermore, it may be important to note that 
in the strongly coupled region JV lattice and JP 
soliton-like wave excitations 
as well as continuous waves are expected as described above.
Further study is needed to obtain better understanding of
these complicated dynamical behaviors.

In summary, we have for the first time found {\it two} microwave excitation modes 
with a temperature-dependent gap in parallel magnetic fields
of BSCCO.
Strong coupling between JP and the JV lattice are 
responsible for these two modes.
We have concluded that in high fields where all block layers are occupied with JVs,
one lying at higher frequencies comes from the JP wave propagating in 
a tilted direction with respect to the JV lattice structure, 
while the other seems to 
originate from the oscillation of the Josephson vortices.

\begin{acknowledgments}
We thank M. Machida, A. Koshelev, T. Koyama, and M. Tachiki
for their fruitful discussions and critical comments.
This work has been supported by CREST, 
the Japan Science and Technology Cooperation.
\end{acknowledgments}

\bibliography{JPJVBibs,MyBibs}

\end{document}